# Opportunistic Delay Tolerant Routing for LED Wristbands in Music Events

Nandita Joshi
*School of Computer Science*
*University of Nottingham*
Nottingham, United Kingdom
psynj1@nottingham.ac.uk

Milena Radenkovic
*School of Computer Science*
*University of Nottingham*
Nottingham, United Kingdom
milena.radenkovic@nottingham.ac.uk

*Abstract*—The existing technology used for LED wristbands does not explore the use of opportunistic delay tolerant network (DTN) routing protocols to disseminate lighting information. This paper multiple trade-offs of different criteria with the protocol (Spray & Focus and PRoPHETv2) changes with varying music event scenario configurations & capacities using the ONE simulator. When considering variable music event durations, it was shown that PRoPHETv2 was more desirable for longer durations however Spray & Focus accounted for network congestion issues that could occur for smaller buffer sizes in the wristbands. When considering variable patterns of audience density, Spray & Focus had higher delivery probabilities whereas PRoPHETv2 had near zero results. Further research can be conducted in incorporating buffer removal management to the protocols as well as experimenting with different congestion targeted DTN protocols.

## I. Introduction

It is found that people are likely to attend 2-5 music events per year regardless of what generation they are from to see their favourite artist [1]. These can range from seeing them at a small gig (capacity within 100 - 1,000), an arena (capacity within 1,000 - 25,000) or a large stadium (capacity 40,000 - 100,000). This means venues throughout the year are filled with fans seeking out the best music experience possible.

Over the years, there have been many improvements made to the experiences of audience members for live music events such as visual effects, production and audience participation. One form of audience participation that is constantly evolving as a technology are LED wristbands/sticks. These allow every audience member that has one to light up in the crowd based on what lighting sequence the artist desires for their set list.

There are three types of technology that currently exist for this. Xyloband uses radio receiver within plastic case in the band and receives these frequencies from a radio transmitter [2]. This typically works best for designated seating areas (no movement) [3]. PixMob uses multiple robotic transmitters that send out lighting data to the wristbands which work regardless of wristband location [4] [5]. BTS Lightsticks use an app that pairs to your seat during a concert, and a lighting team at the venue controls which colour ocean appears in which sections during certain numbers [3].

A new technology can be proposed for LED wristbands that use delay tolerant network (DTN) routing protocols to send messages containing information such as light intensity and colour to the wristbands worn by audience members which act as stationary nodes at events like this. This means the lighting sequence would be less synchronised and preprogrammed, leading to the lighting sequence being based on the routing of the nodes in the network at the time the music is performed. This paper intends to investigate the performance DTN protocols within a concert environment.

## II. Related Work

### A. Peer-to-Peer Networks (P2Ps), Mobile Ad-hoc Networks (MANETs) & Vehicular Ad-hoc Networks (VANETs)

Communication and transfer of data typically occurs through a centralised infrastructure such as an access point through a broadband cellular network transmission towers (4G, 5G, etc) or wireless modem. Three network types, Mobile Ad-hoc Networks (MANETs), Vehicular Ad-hoc Networks (VANETs) and Peer-to-Peer networks (P2P), propose a decentralised approach to node communication that treats nodes as clients, servers and routers at the same time, rather than nodes only being clients (destination nodes) or servers (source nodes).

Peer-to-Peer networks must share resources between one another in order to facilitate services such as file sharing and shared workspaces. P2P networks work in the application layer and do not focus on node mobility. It may be useful to work directly at network level when dealing with mobile nodes [6].

MANETs allow "Ad-hoc" communication within a network of wireless, mobile nodes with a dynamic topology. Due to the limited transmission range of the nodes, the packets are forwarded either directly to the destination node if in range (single hop) or to an intermediate node that has a known path to the destination node (multi-hop). These "known paths" can be computed through different types of MANET routing protocols such as table-driven/proactive (constantly surveys network for all possible routes when network topology changes and stores this in a routing table), reactive (only establish a route to destination when needed thus do not update routing tables constantly) and hybrid (trade-off between proactive and reactive). MANETS are used in a various use cases such as sensor networks, disaster management, military, education and personal area networks [7] [8].

VANETs operate in an almost identical way to MANETs with slight differences. Vehicles act as nodes in the network meaning the scale of the network is much larger than MANETs. Additionally, the node mobility is predictive and high compared to random and low. The communication range of a VANET must be much higher than a MANET as vehicles can pass each other at high speed which gives a shorter connection time thus expanding the range increases the time that the nodes are connection for [9].

## B. Opportunistic Networks & Delay Tolerant Network (DTNs) Routing Protocols

The one limitation of MANETs and VANETs are that they drop packets upon long disconnections which is not ideal due to nodes (wireless devices and vehicles) potentially becoming isolated from other nodes in low network density areas or when nodes suffer loss of power. Opportunistic networks assume that there is no end-to-end path to the destination from the source and supports unpredictable communication patterns. Delay Tolerant Network routing protocols are used for these communication patterns [8].

Delay Tolerant Networks are networks that support long disconnections/delays of nodes. This could occur due to unstable network topology, unstable node channels between nodes or differing node processing capabilities. DTNs follow the store-carry-forward paradigm that all routing protocols must implement. The paradigm overcomes the intermittent connectivity issue by having each node store messages in a buffer, carried to a node that the message can be forwarded to the destination node or a node that can store-carry-forward to the destination node [10]. This can be illustrated in figure 1.

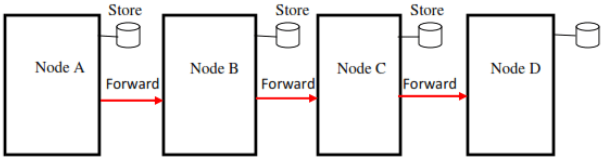

Fig. 1.  The store-carry-forward paradigm visual [11].

Opportunistic networks with DTN routing protocols work in the use case of LED wristbands at concerts as there will not always be a direct route from music source to destination wristband node meaning that intermittent nodes will have to carry the messages. Also, the LED wristbands may be susceptible to loss of power at a given point.

There are two types of DTN routing protocols: forwarding-based and replication-based. Forwarding-based DTN routing protocols focus on reducing the resource/communication overhead of network by generating a single copy of the message to be sent from the source to the destination. This means that the node will find the best path prior to forwarding the message. However, this results in a low message delivery rate as the packet may be dropped due to an unforeseen failure. Replication-based DTN routing protocols are the opposite as they ensure that the message delivery rate of the network is high by sending multiple copies of the same message across the network. This leads to the resource/communication overhead of the network to be higher. There is a time-capacity trade-off between using forwarding-based and replication based DTN protocols. For the LED wristband communication, replication-based protocols will be investigated to maximise the probability that the packet will get to the destination node (the wristbands need to light up).

*1) PRoPHETv2:* Probabilistic Routing Protocol using History of Encounters and Transitivity (PRoPHET) is a replication-based DTN routing protocol that exploits the non-randomness of node mobility to generate delivery predictability metrics that are then used to decide the best sequence of hops from the source node to the destination node, storing these as vectors. The delivery predictability is calculated based on three properties. The first property is to ensure that the metric is updated when a node has been encountered. The second property is to apply an "aging" constant that causes the probability to decay the longer the last encounter was. The last property is to determine the effect of transitivity (how much of an impact predictability metrics from other nodes will have on the current node) [12].

One of the main issues with PRoPHET is that the delivery predictability is updated even with intermittent connections. This means that the protocol can mistake short disconnections and reconnections from the network as new encounters, leading to an overestimate of the number of "real" encounters the node has had which is factored into the transitive property of the delivery predictability calculations. This is called the "Parking Lot Problem".

PRoPHETv2 aims to solve this problem by adjusting the calculation of the delivery predictability to incorporate a metric called the encountered probability which takes into account the last time message exchange occurred with that node and uses that as a factor to reduce the amount that the delivery predictability changes for recently encountered nodes (equation 1 & 2 show the calculation differences of PRoPHET and PRoPHETv2).

$$P_{(a,b)} = P_{(a,b)_{old}} + (1 - P_{(a,b)_{old}}) \times P_{init} \qquad (1)$$

$$P_{(a,b)} = P_{(a,b)_{old}} + (1 - P_{(a,b)_{old}}) \times P_{enc} \qquad (2)$$

Where $P_{enc}$ is calculated by:

$$P_{enc} = P_{max} \times (Intvl_B / I_{typ}); \text{ if } 0 \leq Intvl_B \leq I_{typ} \qquad (3)$$

$$P_{enc} = P_{max}; \text{ otherwise} \qquad (4)$$

PRoPHETv2 also makes aims to solve this problem by changing the transitive predictions by removing the additive impact on the delivery predictability and replacing it with a function that takes the maximum of either the previous predictability metric or the newly calculated transitive impact (see equations 5 & 6).

$$P_{(a,i)} = P_{(a,i)_{old}} + (1 - P_{(a,i)_{old}}) \times P_{(a,b)} \times P_{(b,i)} \times \beta \qquad (5)$$

$$P_{(a,i)} = \max(P_{(a,i)_{old}}, P_{(a,b)} \times P_{(b,i)} \times \beta) \qquad (6)$$

Where *a* and *b* are the nodes considered for the delivery predictability, *i* is an intermediate node that contacts *b*, *β* is the weighting effect of transitivity, $P_{init}$ is the initial predictibility, $Intvl_B$ is time since last encounter with node B and $I_{typ}$ is the expected typical time interval between connections.

With the more accurate delivery predictability, relay nodes will be more effectively selected, leading to better delivery rates and less overhead [13].

*2) Spray & Focus:* A disadvantage of PRoPHET & PRoPHETv2 is that there is no control over how many

replications of the message are sent across the network thus producing high resource/buffer overhead. Spray & Wait allows you to control how many copies of the message are disseminated through the network. This is achieved by conducting two phases. The spray phase acts first by sending L messages from the source node to distinct relay nodes, where L is predetermined by protocol user. The wait phase is enacted if the destination node is not reached in the spray phase. This involves the relay nodes directly forwarding the message to destination node if they encounter it. For a network with static node mobility and a low L value, this could result in a low delivery rate for messages as the TTL (time to live) of the message may expire before the relay node encounters the destination node [14].

Modifications can be made to the spray phase of the protocol to accommodate this. One of these modifications is called Binary Spray & Wait. The spray phase spreads the floor function of the half of the L messages specified to relay nodes, leaving the source node with the ceiling function of half the L messages. The spray phase continues iterating until each node is left with one message which is when the wait phase is enacted (the same as before).

Even with these modifications, the relay node with one message remaining must come in direct contact with the destination node in order for the message to be delivered which is not always the case. Spray & Focus removes the wait phase and replaces it with a focus phase. Similarly to the wait phase, the focus phase is enacted when a node is left with one message. Then, it forwards the message based on a utility function that utilises last encounter times with other nodes to incorporate transitivity into the forwarding decision (similar to PRoPHET & PRoPHETv2). It is forwarded until the destination node is reached (or the TTL expires). Spray & Focus has a binary spray phase version as well [15].

### C. Opportunistic Network Environment Simulator

The Opportunistic Network Environment (ONE) Simulator provides a way to simulate a scenario based around DTN routing protocols and the node mobility based on this routing. It achieves this by allowing the user to modify parameters in a configuration text file that is then loaded into the simulator that can be run for a user specified amount of time. When the simulator is run, the user is able to obtain a visualisation of the scenario at runtime as well as results following the completion of the simulation. The most widely known DTN routing protocols are pre-built into the simulator's source code however it is possible to create/obtain more protocols as they are written/read as Java files. This is the simulator used for modelling the LED wristband use case. There are implementations of PRoPHET and Spray & Focus that can be used on the simulator.

### III. USE OF LED WRISTBANDS IN COMPLEX MUSIC EVENT SCENARIOS

#### A. Design & Implementation

The scenario focuses on LED wristbands within a concert venue as afore identified. In an ideal situation, it would be desired to have all lighting information messages that are sent received as fast as possible due the wristbands needing to be lit up at appropriate time intervals during the event. This can be quantified in the simulator as the delivery probability and latency average. As well as this, a consideration must be made into the amount of resources that are used which can be quantified as overhead in the simulator. Overhead is calculated by finding the number of message relays that occurred that were not to the audience members and dividing that value by the message relays that were to the destination. Using the ONE simulator configuration file, the parameters for the settings can be modified to fit the requirements of the music event scenario to optimise the metrics stated. These settings can be seen in Table I.

TABLE I. ONE SIMULATOR CONFIGURATION SETTINGS

| Configuration | Values |
| --- | --- |
| Network Density | 101, 251, 501 and 1001 nodes |
| Message Size | 1Kb |
| Buffer Size | 1Mb |
| Message TTL | 10 minutes |
| Simulation Time | 1, 2, 3, 4 & 5 hours |
| Communication Interface | Bluetooth |
| Movement Models | GridLocation & StationaryMovement |
| Spray & Focus: Number of Copies | 10, 25, 50 & 100 copies |
| PRoPHETv2: Probability Interval | 10, 25, 50 & 100 seconds |

Within the simulator, nodes can be classified into groups, with each group having different properties to the others. In this scenario, there exists two groups: artist and audience. The artist group act as the source nodes and the audience group act as the destination nodes. Different music events have different crowd capacities, therefore the audience members will be varied (100, 250, 500 & 1000 nodes) with the artist remaining constant (1 node). One set of experiments can test how well the protocols by varying network density.

For this scenario, it was decided that the message size, buffer size and message TTL remain constant as they were not parameters that would change given different music event scenarios. The message size is set to be 1Kb as the data being sent across to the wristbands would consist of light intensity values and RGB values which do not require a large message size to send across. The buffer size is set at 1Mb. The message TTL is set as 10 minutes as it is desired that the lighting information is sent as soon as possible due to the dynamic nature of music played during a concert.

Music event durations can vary depending on type of event it is and what the artist decides for the set list. In the simulator, there is the capability of adjusting the length at which the scenario can last for. In another set of experiments, this duration can be varied for each protocol to see how they perform.

For both the artist and audience members, Bluetooth was chosen as the communication interface. There are many incarnations of Bluetooth. The one used for this scenario is classic Bluetooth as opposed to Bluetooth Low Energy as it is the standard used speakers, headphones and in-car entertainment systems and LED wristbands fall in a similar category to these [16]. The communication range for classic Bluetooth is 10 metres which works in our scenario as audience members will be close to one another (refer to movement model explanation). The transmission speed is set at 250kB/s which is equivalent to 2Mb/s.

Each group in the simulation has a different movement model. The audience group movement model is set to be GridLocation as a concert will either have preset seating at set distances or designated standing areas in a grid formation. The GridLocation movement model rows change based on network density to optimise the perimeter calculation of the audience layout. Additionally, there is a fixed spacing between audience members within the grid at 10 metres as the radius of the communication range is 10 metres which works well.

As well as this, the movement model requires coordinates to specify where on the simulator the audience will generate. The artist group movement model is set to be StationaryMovement which only requires the coordinates of audience members. The configuration file allows for an underlay to the GUI to be specified so seating chart of the O2 arena was chosen for this. The layout & underlay can be seen in figure 2.

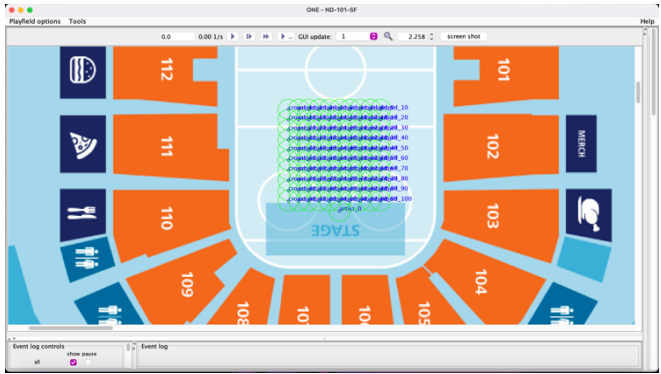

Fig. 2. A screenshot of ONE Simulator with the scenario settings for 100 audience members and 1 artist.

There are protocol specific settings that can be adjusted as experiments to see whether they improve the performance of the outlined metrics. For Spray & Focus, this is the number of copies replicated. For PRoPHETv2, this is how many seconds the aging of delivery predictions are recalculated at.

### B. Protocol Implementation

Unlike Spray & Focus, the PRoPHETv2 router Java file was not previously implemented in the simulator. We implemented a new Java file was created entitled *PRoPHETv2Router.java* that was mostly the same at the *PRoPHETRouter.java* files with some changes.

There were a few minor deviations that were carried out to edit the PRoPHETRouter to fit the functionality. The default $\gamma$ (gamma) and $\beta$ (beta) values were modified from 0.98 and 0.25 to 0.9998 and 0.9 respectively [13]. Two new constants were defined for $P_{enc}$ and $I_{typ}$ with values 0.5 and 1800 respectively [13]. An initialiser for new hash map that stores the last encounter times for all other nodes is called in PRoPHETv2's constructor. A new function called *getEncTimeFor()* which returns the last encounter time (if one exists) with a specific node as seen in the code block:

```java
public double getEncTimeFor(DTNHost host){
    if (lastEncounterTime.
            containsKey(host)) {
        return lastEncounterTime.get(host);
    }
    else {
        return 0;
    }
}
```

One of the two major changes was to the updateDeliveryPredFor() function where we are implementing equation 2 (including calculating the encountered node probability) from the PRoPHETv2 paper [13]. The modified code is shown:

```java
private void updateDeliveryPredFor(DTNHost
        host) {
    double PEnc;
    double simTime = SimClock.getTime();
    double lastEncTime=getEncTimeFor(host);

    if(lastEncTime==0)
        PEnc=PEncMax;
    else {
        if((simTime-lastEncTime)<I_TYP){
            PEnc=PEncMax*((simTime-
                lastEncTime)/I_TYP);
        }
        else
            PEnc=PEncMax;
    }

    double oldValue = getPredFor(host);
    double newValue = oldValue +
        (1 - oldValue) * PEnc;
    preds.put(host, newValue);
    lastEncounterTime.put(host, simTime);
}
```

The second change was to the *updateTransitivePreds()* function which, similar to the previous function, involves using the new transitivity calculation (equation 4) as outlined in the PRoPHETv2 paper [13]. Most of this function stayed the same, the only line that needed to be modified to accommodate these:

```java
private void updateTransitivePreds(DTNHost
        host) {
    MessageRouter otherRouter = host
        .getRouter();
    assert otherRouter instanceof
        ProphetV2Router :
        "PRoPHETv2 only works with other
        routers of same type";

    // P(a,b)
    double pForHost = getPredFor(host);
    Map<DTNHost, Double> othersPreds =
        ((ProphetV2Router)otherRouter)
        .getDeliveryPreds();

    for (Map.Entry<DTNHost, Double>
        e : othersPreds.entrySet()) {
        if (e.getKey() == getHost())
            continue;

        //ProphetV2 max(old,new)
        double pOld = getPredFor(e
                    .getKey());
        double pNew = pForHost *
            e.getValue() * beta;

        if(pNew>pOld)
            preds.put(e.getKey(), pNew);

    }
}
```

## IV. MULTI-DIMENSIONAL PROTOCOL PERFORMANCE ANALYSIS

### A. Dynamic Music Event Duration

When investigating simulation duration, the protocol parameters (number of copies = 50, probability interval = 50) and network density (100 audience members and 1 artist) remained constant.

When investigating the delivery probability of both protocols, Spray & Focus and PRoPHETv2 (at some point) experience at gradual decrease in delivery probability as time goes on (see Figure 3). For Spray & Wait, there is a consistent decrease in delivery probability with time which could be caused by the buffers of the audience members closest to the artist and the artist itself filling up. This is because the binary mode of Spray & Focus will always send (in our case) 25 messages to the static neighbouring audience members and keep 25 in its own buffer which builds up over time. Subsequently, this leads to the message TTL to expire thus dropping the message, lowering the delivery probability. For PRoPHETv2, the delivery probability mostly remains at a constant of 1, as for a static network, encountered audience members remain the same for each of them meaning the delivery predictability metrics remain the same. Thus, an optimal sequence of hops from the artist to varying audience members can be found with ease. The sharp decrease at 5 hours could be caused by buffer overflows similar to Spray & Wait but this time, it would impact the delivery predictability of next hops. If an audience member hasn't delivered a lighting information in a while due the buffer, the delivery predictability would decrease for that audience member and others would not choose to send lighting information to it even though it is the optimal path. By taking the less optimal path, the message TTL might expire thus dropping the message and lowering the delivery probability.

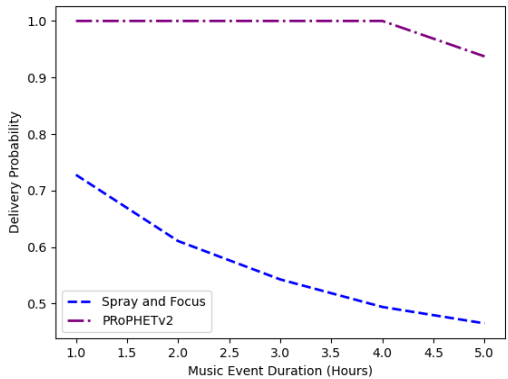

Fig. 3. A graph of how the delivery probability changes given the music event duration for each protocol.

When investigating the overhead and latency of both protocols, it is shown that the overhead and latency of PRoPHETv2 are substantially higher than Spray & Focus (see Figure 4). For PRoPHETv2, the overhead and latency remain at a constant value until the 5-hour mark where we see a large spike. This resonates with the results from the delivery probability which suggests that the same reasoning of why this occurs applies here too (buffers reaching capacity). For Spray & Focus, the overhead gradually increases at a constant rate. This means that, over time, more messages are being relayed than being delivered which can be due to wristband buffer capacities filling up causing messages to take different routes to the audience members. The latency average for Spray & Focus gradually increases between 1-4 hours but spikes at 5 hours. That spike could be due to almost all buffer capacities being full and being held at a standstill until some messages are delivered to the desired audience member. More generally, in the instance of a network of static nodes, network topology has an impact on latency.

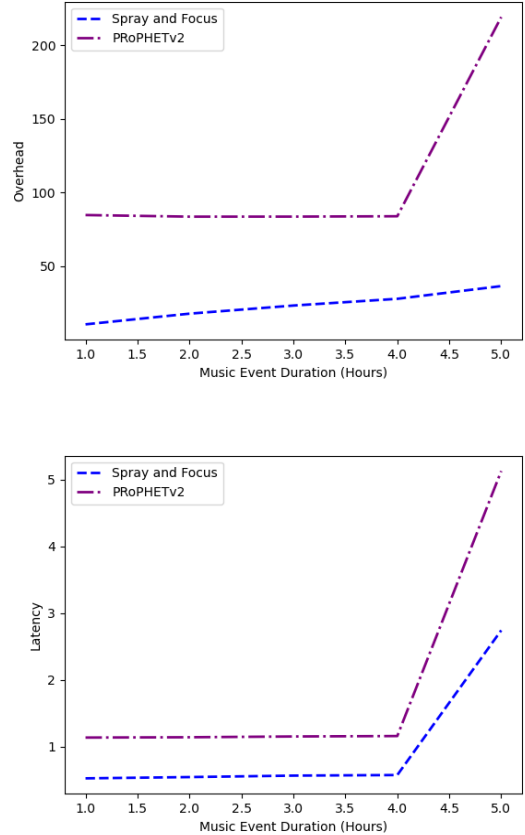

Fig. 4. Graphs of how the overhead and latency average changes given the music event duration for each protocol (top to bottom).

Overall, PRoPHETv2 seems desirable when running longer simulations due to its consistent and high delivery probabilities. Whilst the overhead and latency averages are higher than Spray & Focus, they are mostly consistent which can be accommodated when considering finite resources. However, in situations where buffer size may have to be reduced for the LED wristbands, it might be worth using Spray & Focus to avoid large scale network congestion, as seen at the 5-hour mark.

### B. Dynamic Audience Density

When investigating audience density, the protocol parameters (number of copies = 50, probability interval = 50) and music event duration (2 hours) remained constant.

When investigating the delivery probability of both protocols, whilst PRoPHETv2 starts off with a higher probability for 100 audience members than Spray & Focus, PRoPHETv2's delivery probability drastically decreases to zero and Spray & Focus's delivery probability increases (see Figure 5). PRoPHETv2's descent to zero could be due to the message TTL expiring before any lighting information messages could be delivered due to how large the networks became. Alternatively, it could be caused by the protocol not being able to compute all delivery predictabilities accurately and in time for lighting information messages to be forwarded. This would leave messages being forwarded to the wrong places or remaining in the artist source node buffer. Spray & Focus would thrive in high density networks as it means it has access to more buffers as well as more paths to the audience members. This implies that it is less likely that the message TTL will expire before the message has been delivered.

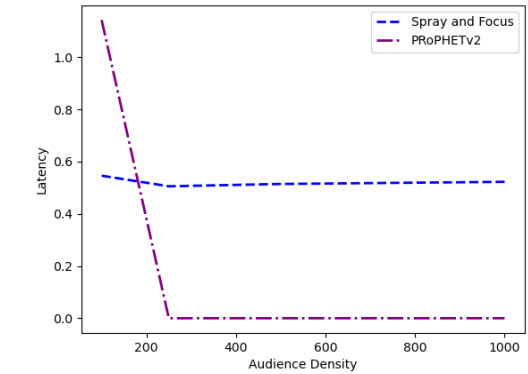

Fig. 6. A graph of how much overhead is generated and the latency average given the audience density for each protocol (top to bottom).

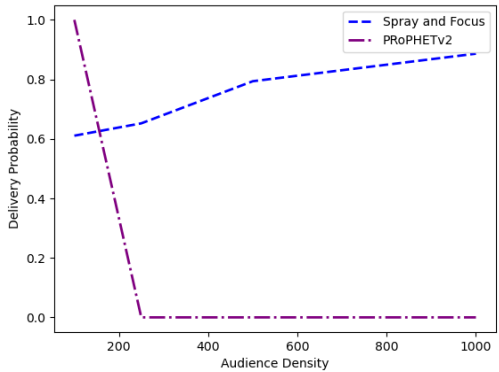

Fig. 5. A graph of how the delivery probability changes given the audience density for each protocol.

Both protocols follow the same trend in that the higher the audience density, the lower/constant the overhead and latency average becomes (see Figure 6). PRoPHETv2's results for both metrics show high values for audience density of 100 and then a sharp decrease to zero for all other network densities experimented. Considering that the delivery probabilities were zero for PRoPHETv2 for every network density other than 101 as well, this could mean that the messages did not leave the buffer of the artist source node. This could be due to the fact that the delivery predictability deems none of audience are within range of the artist to send the lighting information so keeps then it. The network mobility is static which means that the artist will never encounter an audience member that it deems appropriate to send the message, leading to no relays and subsequently no messages delivered. Spray & Focus overhead also becomes zero for every audience density experimented other than 100. A zero value for overhead and non-zero values for delivery probability suggests that the only message relays that occur are ones directly to the audience members from the artist. This means that only audience members that are in the communication range of the artist are getting lighting information delivered to them.

Both protocols show the same shape of graph (following a 1/n relationship) for average number of messages per node (just a different y-intercepts) but follow different trends for number of messages created (see Figure 7). The average number of messages per node was calculated by taking the number of messages created and dividing it by the audience density. This is useful to know in our scenario as it is desired to have as many lighting messages sent to each audience member as possible so that they experience more changes on their wristbands. Spray & Focus produces far more messages than PRoPHETv2 for the same reasons as the delivery probability results (see above). For the average number of messages per audience member, the shape of the graphs suggest that the Spray & Focus average will tend to 4.5 and the PRoPHETv2 average will tend to 0.5. The gradual decrease for Spray & Focus could be explained by the message TTL expiring before the message even can reach the desired audience member as it would have more hops to complete for audience members that are not bottom centrally located (close to the artist). It could be worth, in further research, to investigate this message TTL for different audience densities. The gradual decrease for PRoPHETv2 can be explained by the created messages not increasing due to messages not leaving source node buffer (same reasoning as for PRoPHETv2 delivery probability results) but the audience density continuing to increase. If messages were not only being delivered to the audience members closest to the artist, Spray & Wait would better to send more lighting information to the LED wristbands in high audience density music events.

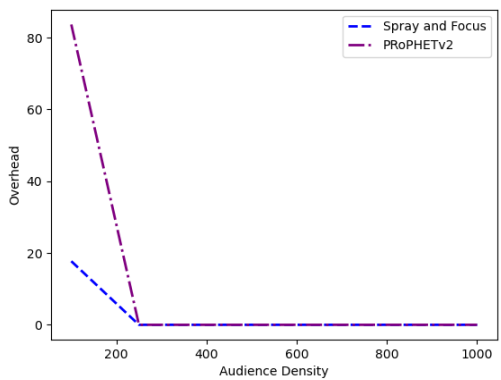

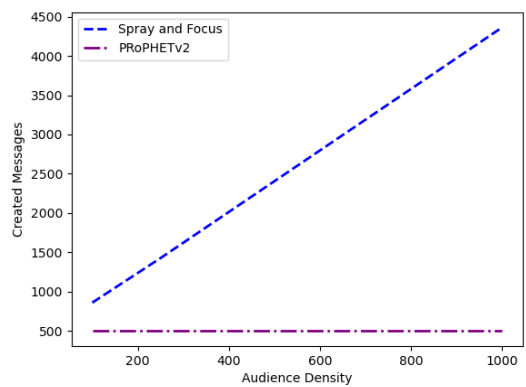

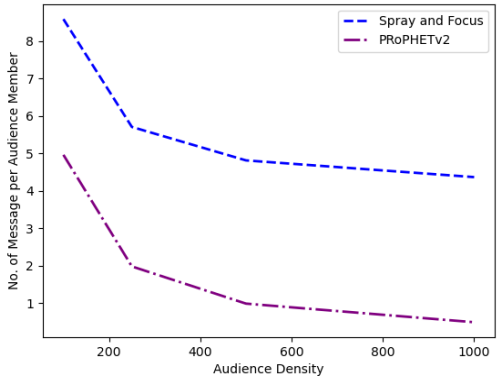

Fig. 7. Graphs of how the created messages and messages per destination node metrics change given the audience density for each protocol (top to bottom).

Considering that PRoPHETv2's delivery probability, latency average and overhead were zero for all but one network density, it could suggest that the ONE simulator could not handle the protocol being run for high network densities. Equally, it could suggest that the protocol cannot handle high network densities. In contrast, Spray & Focus does have higher delivery probabilities and creates more messages however, it is suggested that the only audience members receiving the messages are ones closest to the artist. Subsequently, both do not fare well for increasing network density.

*Note: An audience density of 10,000 was attempted, but the simulator ran tremendously slow and took too long to produce meaningful results. This would have been useful as this is a typical music event capacity but could not be achieved.*

### C. Varying Protocol Parameters

When investigating protocol specific parameters, the network density (100 audience members and 1 artist) and music event duration (2 hours) remained constant.

When the protocol parameters for Spray & Focus and PRoPHETv2 were increased, the delivery probability was either improved or remained constant (see Figure 8). Spray & Focus sees a gradual increase in delivery probability. By increasing the number of copies, it allows the message to have more opportunities via different audience member hops to propagate through the crowd faster, reaching the desired audience member before the message TTL expires. PRoPHETv2 consistently stays at a delivery probability of 1 which means that given the change in predictability calculation intervals, the messages are still delivered at, what the protocol believes to be, the optimal path of hops.

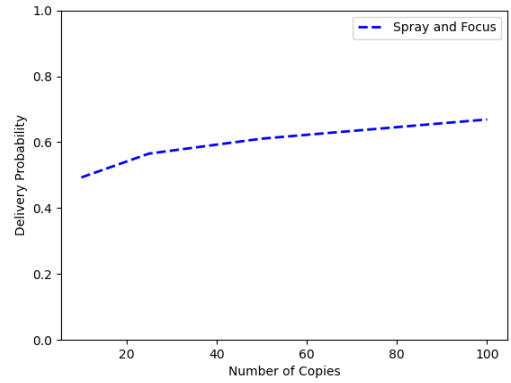

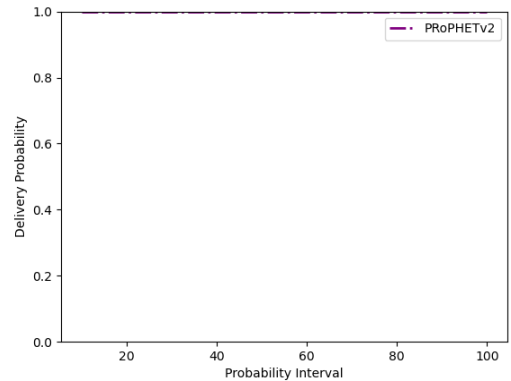

Fig. 8. Graphs of how the delivery probability changes given the parameter changes for each protocol (top to bottom).

Both PRoPHETv2 and Spray & Focus showed performance to have improved or stay consistently good when their protocol parameters are increased. Spray & Focus sees a higher value of number of copies benefiting the message delivery probability however there is a trade-off as the higher the number of copies is, a higher buffer capacity is needed to support this. In the case of the concert scenario, this resource is finite. When picking probability interval for PRoPHETv2, it is desired to pick a higher interval without the delivery probability decreasing as it means that the tables that each audience member has for the predictability for all other audience members are refreshed at a slower rate meaning less calculations are completed overall.

### V. WIDER DISCUSSION OF RESULTS

The Multi-Dimensional Protocol Performance Analysis section shows that there is not one protocol that performs better overall for the concert scenario than the other as there are advantages and disadvantages of both. It was shown that PRoPHETv2 works more consistently when considering delivery probabilities for long simulation durations (which in our scenario would be the duration of a concert). One trade off to consider with PRoPHETv2 in this scenario would be that resources are limited. This means that there are limitations on buffer capacities and number of hops that can be carried out by each audience member. The more resources used, the more battery for the wristbands is consumed. On the other hand, for high network densities (larger concert capacities), both protocols seem to falter due to buffers reaching maximum capacity.

Mashhadi et al discuss the implications of fair content dissemination in DTNs surrounding given the scenario of local community users (festivals, sports races and parties). This aims to offer a reduction in overhead whilst also maintaining a high delivery probability by adding a load balancing layer (prevention and alleviation) on top of a DTN protocol. This load balancing was achieved by making each node monitor the amount of traffic it has been forwarded and if it reaches a limit, the node will back off from the network and rejoin when it has less of a message load. The results show that this novel approach does improve delivery probabilities whilst reducing overhead [17].

As given above (Figure 4), both PRoPHETv2 and Spray & Focus generate a large resource overhead when simulation duration is increased. By applying load balancing to the scenario presented in this paper, a minimisation of this overhead can be achieved whilst maintaining the high delivery probability rather than having to trade off performance and resource usage.

It was required that the nodes deliver, on average, a large amount of messages regarding the lighting settings to the wristbands to create a sense of continual audience participation however, this was seen to a high level for both protocols with Spray & Focus sending on average less than 9 messages to each audience member in 2 hours and PRoPHETv2 sending on average less than 5 messages to each audience member in 2 hours. Non DTN-based approaches for LED wristband technology perform much better than this, with changes to lighting settings of each wristband happening almost every second.

Even though Opportunistic Networks and Delay Tolerant Networks do not seem better than what currently exists for LED wristband technology, there are scenarios where the results presented show promise for DTNs being implemented in different use cases. Radenkovic et al describe a technique that allows DTN protocols for a VANET to be used to facilitate vehicular charging (CognitiveCharge protocol). It achieves this by monitoring how quickly the battery is being used up over time and based on that metric, decides whether to share, obtain or withhold their charge. It also makes use of an Ego Network which holds information on nodes that are not itself and their relationships with the other nodes in the network [18].

The results of this paper were found by experimenting with the CognitiveCharge protocol within a VANET of electric vehicles (EVs) and stationary points of interest which would be EV charging points. It was experimented in two scenarios Nottingham and San Francisco. In both cases, the results showed a decrease in vehicles below critical battery threshold and a reduced wait time to charge compared to not using a protocol at all. This shows that there are relevant scenarios in which DTN protocols would be useful.

## VI. FUTURE WORK

As discussed briefly in both the Multi-Dimensional Protocol Performance Analysis and Wider Discussion of Results sections, there is further research that can be conducted for this scenario to see if improvements can be made to the performance of DTN protocols in LED wristbands so that they can operate as well as other technologies used for them. In addition to this, different variations of the scenario can still be tested.

One of the biggest issues found in the network density for both the Spray & Focus and PRoPHETv2 experimentations was that the buffers of the audience members wristbands in the network would reach maximum capacity from messages that had either already been delivered or had been dropped. Currently, there is no mechanism to remove these messages from the buffer in the experiment. Abbasi et al explores a novel approach to dropping messages in node buffers. It outlines multiple methodologies for policies that can be implemented alongside the protocols that have network knowledge [19]. For PRoPHETv2, it could utilise an Evict Least Probable First (LEPR) which would evict messages that collectively have low delivery predictabilities for neighbouring nodes. For Spray & Focus, it could utilise Evict Most Forwarded First (MOFO) to drop excess copies of the messages from the protocol that never reached destination node. By adding these policies to the experiments, it's impact on delivery probability, overhead and latency average can be investigated.

Instead of adding a buffer management policy alongside a protocol, a new protocol investigating the scenario could be proposed - specifically, replication management-based protocols. CafREP is a protocol that uses social, resource and regional driven heuristics to forward messages, aiming to minimise congestion. Based on a comparison, the protocol decides whether to forward a message and how many copies of the message to forward [20]. The audience density and music event duration can be varied to test how the protocol works in different concert scenarios.

Another protocol consideration that can be made is the use of routing lighting information messages using deep learning techniques. Radenkovic et al explore CognitiveCache, "a multi-agent deep reinforcement learning approach" to counteract duplicate message traffic and network congestion [21]. This could aid the issues that arose with audience density increases.

Scenario based adjustments can be made to test for different music event cases. An experiment could be made to account for the fact that there could be multiple artists on stage acting as source nodes, accounting for what the delivery probability, latency average and overhead would be as artists increase. Also, festival music events and areas of concert music events can have audience members not seated (not stationary) which means the movement model of them would need to change in the experiment settings. Then, all simulations varying audience density and music event duration could be rerun to see whether the results vary dramatically.

It is important that the music event scenarios outlined within the paper can be deployed in the real world using embedded systems/systems on a chip (SoC). There is emerging work of real-world deployments of MODiToNeS and personal edge cloud smart IOT devices (RasPiPCloud) in different contexts such as smart cities, smart rural, smart manufacturing and smart healthcare scenarios [22][23][24]. We envisage that we will deploy our music event routing protocols using different smart edge fully autonomous devices that can sense, learn, predict and communicate with each other such as smart wristbands that would be controlling LED lighting messages in venues/festivals.

## REFERENCES

[1] H. Schmidt, "Concert attendance and enthusiasm: A survey," Sep 2023. [Online]. Available: https://www.innerbody.com/ concert-attendance-and-enthusiasm#

[2] Wikipedia, "Xyloband" Oct 2023. [Online]. Available: https://en.wikipedia.org/wiki/Xyloband

[3] D. Conradie, "Exploring the tech behind concert led wristbands," Jul 2023. [Online]. Available: https://hackaday.com/2023/07/01/exploring-the-tech-behind-concert-led-wristbands/

[4] PixMob, "Led wristbands for enhanced crowd experiences." [Online]. Available: https://www.pixmob.com/products/detail/led-wristbands

[5] Y. Li, "The tech behind taylor swift concert wristbands," Oct 2023. [Online]. Available: https://wired.me/technology/ the- tech- behind-taylor-swift-concert-wristbands/

[6] [6] M. Castro, A. Kassler, C.-F. Chiasserini, C. Casetti, and I. Korpeoglu, *Peer-to-Peer Overlay in Mobile Ad-hoc Networks*, 01 2010, pp. 1045–1080.

[7] J. Hoebeke, I. Moerman, B. Dhoedt, and P. Demeester, "An overview of mobile ad hoc networks: Applications and challenges," *JOURNAL-COMMUNICATIONS NETWORK*, vol. 3, pp. 60–66, 07 2004.

[8] A. Bhagat, R. Chaudhari, and K. Dongre, "Content-based file sharing in peer-to-peer networks using threshold," *Procedia Computer Science*, vol. 79, pp. 53–60, 12 2016.

[9] N. Benamar, K. D. Singh, M. Benamar, D. El Ouadghiri, and J.-M. Bonnin, "Routing protocols in vehicular delay tolerant networks: A comprehensive survey," *Computer Communications*, vol. 48, pp. 141–158, 2014, opportunistic networks. [Online]. Available: https://www.sciencedirect.com/science/article/pii/S0140366414001212

[10] A. V. Vasilakos, Y. Zhang, and T. Spyropoulos, *Delay Tolerant Networks: Protocols and Applications*, 1st ed. USA: CRC Press, Inc., 2011.

[11] U. Malekar and L. Kulkarni, Analyzing Credit Based Incentive Mechanisms in Delay Tolerant Network: A Survey, 01 2015.

[12] A. Lindgren, A. Doria, and O. Schelén, "Probabilistic routing in intermittently connected networks," *SIGMOBILE Mob. Comput. Commun. Rev.*, vol. 7, no. 3, p. 19–20, jul 2003. [Online]. Available: https://doi.org/10.1145/961268.961272

[13] S. Grasic, E. Davies, A. Lindgren, and A. Doria, "The evolution of a dtn routing protocol - prophetv2," in *Proceedings of the 6th ACM Workshop on Challenged Networks*, ser. CHANTS '11. New York, NY, USA: Association for Computing Machinery, 2011, p. 27–30. [Online]. Available: https://doi.org/10.1145/2030652.2030661

[14] T. Spyropoulos, K. Psounis, and C. S. Raghavendra, "Spray and wait: An efficient routing scheme for intermittently connected mobile networks," in *Proceedings of the 2005 ACM SIGCOMM Workshop on Delay-Tolerant Networking*, ser. WDTN '05. New York, NY, USA: Association for Computing Machinery, 2005, p. 252–259. [Online]. Available: https://doi.org/10.1145/1080139.1080143

[15] T. Spyropoulos, K. Psounis and C. S. Raghavendra, "Spray and focus: Efficient mobility-assisted routing for heterogeneous and correlated mobility," in *Fifth Annual IEEE International Conference on Pervasive Computing and Communications Workshops (PerComW'07)*, 2007, pp. 79–85.

[16] "Bluetooth technical overview." [Online]. Available: https://www.bluetooth.com/learn-about-bluetooth/tech-overview/

[17] A. J. Mashhadi, S. B. Mokhtar, and L. Capra, "Fair content dissemination in participatory dtns," *Ad Hoc Networks*, vol. 10, no. 8, pp. 1633–1645, 2012, special Issue on Social-Based Routing in Mobile and Delay-Tolerant Networks. [Online]. Available: https://www.sciencedirect.com/science/article/pii/S1570870511001259

[18] M. Radenkovic and A. Walker, "Cognitivecharge: Disconnection tolerant adaptive collaborative and predictive vehicular charging," in *Proceedings of the 4th ACM MobiHoc Workshop on Experiences with the Design and Implementation of Smart Objects*, ser. SMARTOBJECTS '18. New York, NY, USA: Association for Computing Machinery, 2018. [Online]. Available: https://doi.org/10.1145/3213299.3213301

[19] S. Khan, O. u. Rehman, I. A. Abbasi, H. Hashem, K. Saeed, M. F. Majeed, and S. Ali, "Ss-drop: A novel message drop policy to enhance buffer management in delay tolerant networks," *Wireless Communications and Mobile Computing*, vol. 2021, p. 9773402, 2021. [Online]. Available: https://doi.org/10.1155/2021/9773402

[20] M. Radenkovic and A. Grundy, "Efficient and adaptive congestion control for heterogeneous delay-tolerant networks," *Ad Hoc Networks*, vol. 10, no. 7, pp. 1322–1345, 2012.

[21] M. Radenkovic and V. S. H. Huynh, "Cognitive caching at the edges for mobile social community networks: A multi-agent deep reinforcement learning approach," *IEEE Access*, vol. 8, pp. 179 561–179 574, 2020.

[22] M. Radenkovic, V. S. Ha Huynh, R. John, and P. Manzoni, "Enabling real-time communications and services in heterogeneous networks of drones and vehicles," in *2019 International Conference on Wireless and Mobile Computing, Networking and Communications (WiMob)*, 2019, pp. 1–6.

[23] M. Radenkovic, J. Crowcroft, and M. H. Rehmani, "Towards low cost prototyping of mobile opportunistic disconnection tolerant networks and systems," *IEEE Access*, vol. 4, pp. 5309–5321, 2016.

[24] M. Radenkovic and N. Milic-Frayling, "Demo: RasPiPCloud: A lightweight mobile personal cloud," in *Proceedings of the 10th ACM MobiCom Workshop on Challenged Networks*, ser. CHANTS '15. New York, NY, USA: Association for Computing Machinery, 2015, p. 57–58. [Online]. Available: https://doi.org/10.1145/2799371.2799373